\documentstyle[11pt]{article}
\input epsf
\begin{document}

\thispagestyle{empty}
\begin{center}

\medskip
\vspace{2.5 cm} {\Large \textbf{Integrable KdV Hierarchies \\On $T^2=S^1\times S^1$}}\\
\vspace{2 cm}\textbf{M.B. SEDRA}\footnote{Associate
of ICTP: sedra@ictp.it}\\
{\small \ Universit\'{e} Ibn Tofail, Facult\'{e} des
Sciences, D\'{e}partement de Physique,}\\

{\small \ \ \ }{\small \ Laboratoire de Physique de La Mati\`ere
et Rayonnement (LPMR), K\'{e}nitra, Morocco}\\

{\small \ Groupement National de Physique de Hautes Energies,
GNPHE, Morocco,}\\

{\small \ Abdus Salam International Centre for Theoretical
Physics, Trieste, Italy.}\\

\end{center}
\vspace{0.5cm} \centerline{\bf Abstract} \baselineskip=18pt
\bigskip
Following our previous works on extended higher spin symmetries on
the torus we focus in the present contribution to make a setup of
the integrable KdV hierarchies on $T^2=S^1\times S^1$. Actually two
particular systems are considered, namely the KdV and the Burgers
non linear integrable model associated to currents of conformal
weights $(2,2)$ and $(1, 1)$ respectively. One key steps towards
proving the integrability of these systems is to find their Lax pair
operators. This is explicitly done and a mapping between the two
systems is discussed.
\\

\hoffset=-1cm \textwidth=11,5cm \vspace*{1cm}
\hoffset=-1cm\textwidth=11,5cm \vspace*{0.5cm}
\newpage

\section{Introduction}

Integrable systems \cite{1,2,3,4,5,6} deal with nonlinear differential
equations that we can solve explicitly or by using the inverse scattering
method based on the Lax formulation \cite{1, 2}. The particularity of $2d$
integrable systems is due, in on hand, to the pioneering role that they
deserve to the KdV differential equation and on the other hand to the strong
connection existing with conformal symmetries \cite{CFT} and their higher
spin extensions \cite{zamo1, zamo2}. Since much more spectacular
realizations are done for the integrability of KdV hierarchies in the diff$%
(S^{1})$ case, we focus in this work to study some properties related to KdV
hierarchies in the diff$(T^{2})$ case. This is motivated, in one hand, by
the increasingly important role that play integrable systems and higher spin
symmetries in many areas of physics and mathematics. The best known examples
are given by the Virasoro algebra, which underlies the physics of $2d$
conformal field theories (CFT) and its $W_{k}$-extensions. On the other
hand, it's today well recognized that $2d$ conformal symmetry and it's $W_{k}
$ higher spin extensions are intimately related to the algebra of diff$%
(S^{1})$ and diff$(T^{2})$ respectively \cite{frap, anton, sss, sed07}. In
this context, and after a setup of our conventional notations and basic
definitions, we develop a systematic analysis leading to an explicit
derivation of the KdV and Burgers differential equations. These systems are
based on particular diff$(T^{2})$-non standard Lax operators $%
\{logH,.\}^{2}+u_{2}$ and $\{logH,.\}+u_{1}$ respectively and where the
hamiltonian vector field $\xi _{H}\equiv \{logH,\}$ plays the role of the
derivation in diff$(T^{2})$.

Among the results of this study the possibility to connect these systems,
whose fields $u_{k}$ are living on the bidimensional torus $T^{2}$, through
a consistent mapping that we will setup. Several important properties are
discussed with some concluding remarks at the end.

\section{Setup of the KdV integrable Hierarchy}

\subsection{Diff$(T^2)$: Basic properties}

In this section we give the general setting of the basic properties of the
algebra of bianalytic fields defined on the bidimensional torus $T^{2}$.
\newline
{\bf 1)} The two dimensional torus $T^{2}$ is viewed as a submanifold of the
$4d$ real space ${\bf R}^{4}\approx {\bf C}^{2}$ parametrized by two
independent complex variables $z$ and $\omega $ and their conjugates $\bar{z}
$ and $\bar{\omega}$ satisfying the constraint equation ${z\bar{z}}=\omega
\bar{\omega}=1$. Solutions of these equations are given by $z=e^{in\theta
},\omega =e^{im\psi }$ where $n$ and $m$ are two integers and where $\theta $
and $\psi $ are two real parameters.\newline
{\bf 2)} We identify the ring ${\cal R}$ of bianalytic fields on $T^{2}$
with ${\cal R}\equiv \widehat{\Sigma }^{(0,0)}$ the tensor algebra of
bianalytic fields of arbitrary conformal spin. This is a an infinite
dimensional $SO(4)$ Lorentz representation that can be written as
\begin{equation}
\widehat{\Sigma }^{(0,0)}=\oplus _{k\in Z}\widehat{\Sigma }_{(k,k)}^{(0,0)}
\end{equation}%
where the $\widehat{\Sigma }_{(k,k)}^{(0,0)}$'s are one dimensional $SO(4)$
irreducible modules corresponding to functions of bianalytic conformal spin $%
(k,k)$. The generators of $\widehat{\Sigma }_{(k,l)}^{(0,0)}$ are biperiodic
arbitrary functions that we generally indicate by $f(z,\omega )$ given by
\begin{equation}
f(z,\omega )=\sum\limits_{n,m\in Z}f_{nm}z^{n}\omega ^{m},\hspace{1cm}%
\partial _{\bar{z}}f=\partial _{\bar{\omega}}f=0
\end{equation}%
where the constants $f_{nm}$ are the Fourier modes of $f$. This is nothing
but a generalization of the usual Laurent expansion of conformal fields on
the complex plane ${\bf C}$. Note by the way that the integers $n$ and $m$
carried by the Fourier modes $f_{nm}$ are nothing but the $U(1)\times U(1)$
Cartan charges of the $SO(4)\approx SU(2)\times SU(2)$ Lorentz group of the
Euclidan space ${\bf R}^{4}$. Bianalytic functions on ${\bf C}^{2}$ carrying
$U(1)\times U(1)$ charges $r$ and $s$ and generalizing eq.() are given by
\begin{equation}
f_{{(r,s)}}(z,\omega )=\sum\limits_{n,m\in Z}f_{nm}z^{n-r}\omega ^{m-s};,%
\hspace{1cm}r,s\in Z,
\end{equation}%
The coefficients $f_{nm}$ are given by
\begin{equation}
f_{mn}=\oint\limits_{c_{1}}\frac{dz}{2i\pi }z^{-n-l+r}\oint\limits_{c_{2}}%
\frac{d\omega }{2i\pi }\omega ^{-m+-l+s}f_{(r,s)}(z,\omega ),
\end{equation}%
where $c_{1}\times c_{2}$ is the contour surrounding the singularity $%
(z,\omega )=(0,0)$ in the complex space. \newline
{\bf 3)} The special subset $\widehat{\Sigma }_{(k,k)}^{(0,0)}\subset {\cal R%
}$ is generated by bianalytic functions $f_{(k,k)}$, $k\geq 2$. They can be
thought of as the higher spin currents involved in the construction of the $%
W $-algebra on $T^{2}$ \cite{sss}. As an example, the following fields
\begin{equation}
\begin{array}{lcl}
W_{(2,2)} & = & {u}_{(2,2)}(z,\omega ) \\
W_{(3,3)} & = & {u}_{(3,3)}(z,\omega )-\frac{1}{2}\{logH,{u}_{(2,2)}\}%
\end{array}%
\end{equation}%
are shown to play the same role of the spin-$2$ and spin-$3$ conserved
currents of the Zamolodchikov $W_{3}$ algebra \cite{zamo1, zamo2}. Next we
will denote, for simplicity, the fields $u_{(k,k)}(z,\omega )$ of conformal
spin $(k,k)$, $k\in Z$ simply as $u_{k}(z,\omega )$.\newline
{\bf 4)} The Poisson bracket on $T^{2}$ is defined as follows
\begin{equation}
\{f,g\}=\partial _{z}f\partial _{\omega }g-\partial _{z}g\partial _{\omega }f
\end{equation}%
with $\{z,\omega \}=1$. We denote $\{f,.\}=\xi _{f}$ and $\xi
_{f}.g=\{f,.\}.g=\{f,g\}+g\{f,.\}$ equivalently this shows how the Poisson
bracket on the torus can play the role of a derivation. For convenience we
will adopt the following notation $\xi _{H}\equiv \xi _{logH}$ as been the
hamiltonian vector field operator associated to the arbitrary function $H$.
\newline
{\bf 5)} We present here bellow the essential properties of the objects
involved in this study
\begin{equation}
\begin{tabular}{cc}
{\bf Objects ${\cal O}$} & \hspace{1cm} {\bf The conformal weight $|{\cal O}%
| $} \\
&  \\
$z$, $\omega $, $\partial _{z}$, $\partial _{\omega }$ & $|z|=(-1,0)$, $%
|\omega |=(0,-1)$, $|\partial _{z}|=(1,0)$, $|\partial _{\omega }|=(0,1)$ \\
$L_{k,l}$ & $|L_{k,l}|=(-k,-l)$ \\
$W_{s}(z,\omega ),s=2,3,...$ & $|W_{s}(z,\omega )|=(s,s)$ \\
$\{f,g\}^{(k)}=\underbrace{\{f,\{f,{...}\{f}\limits_{k},g\}$ & \hspace{0cm} $%
|\{f,g\}^{(k)}|=(k,k)+k|f|+|g|$ \\
$\{f,g\}^{k}=\underbrace{\{f,g\}{...}\{f,g\}}\limits_{k}$ & \hspace{0cm} $%
|\{f,g\}^{k}|=(k,k)+k|f|+k|g|$ \\
$\xi _{H}=\{logH,.\}$ & \hspace{0cm} $|\xi _{H}|=(1,1)$ \\
$Res$ & $|Res|=(1,1)$%
\end{tabular}%
\end{equation}

\subsection{The space ${\widehat \Sigma}_{n}^{(r,s)}$ and conformal symmetry}

To start let's precise that the space ${\widehat{\Sigma }}_{n}^{(r,s)}$
contains differential operators of fixed conformal spin $(n,n)$ and degrees
(r,s), type
\begin{equation}
{\cal L}_{n}^{(r,s)}(u)=\sum_{i=r}^{s}u_{n-i}(z,\omega )\circ \xi _{H}^{i},
\end{equation}%
These are $\xi _{H}$'s polynomial differential operators extending the
hamiltonian field $\xi _{H}=\{logH,\}$. Elements ${\cal L}_{n}^{(r,s)}(u)$
of ${\widehat{\Sigma }}_{n}^{(r,s)}$ are a generalization to $T^{2}$ of the
well known KdV operator $\partial _{z}^{2}+u_{2}(z)$. Moreover, eq.(8) which
is well defined for $s\geq r\geq 0$ may be extended to negative integers by
introducing pseudo-differential operators of the type $\xi _{H}^{-k}$, $k>1$%
, whose action on the fields $u_{s}(z,\omega )$ is given by the Leibnitz
rule. Striking resemblance with the standard case \cite{6} leads us to write
the following Leibnitz rules
\begin{equation}
\xi _{H}^{n}{f}(z,\omega )=\sum_{s=0}^{n}c_{n}^{s}\{logH,{f}\}^{(s)}\xi
_{H}^{n-s},
\end{equation}%
and
\begin{equation}
\xi _{H}^{-n}{f}(z,\omega )=\sum_{s=0}^{\infty }(-)^{s}c_{n+s-1}^{s}\{logH,{f%
}\}^{(s)}\xi _{H}^{-n-s}
\end{equation}%
where the $k^{th}$-order derivative $\{logH,{f}\}^{(k)}=\underbrace{%
\{logH,\{logH,{...}\{logH,}\limits_{k\hspace{0.1cm}times}f\}...\}\}$ on the
torus $T^{2}$ is the analogue of $f^{(k)}=\frac{\partial ^{k}f}{\partial
z^{k}}$, the ${k}$th derivative of $f$ in the standard case. The algebra $%
sl_{n}-\widehat{\Sigma }_{n}^{(0,n)}$ describes simply the coset space ${%
\widehat{\Sigma }}_{n}^{(0,n)}/{\widehat{\Sigma }}_{n}^{(1,1)}$ of $sl_{n}$%
-Lax operators on the torus $T^{2}$ given by
\begin{equation}
{{}{\cal L}_{n}}(u)=\xi _{H}^{n}+\sum_{i=0}^{n-2}u_{n-i}\xi _{H}^{i}
\end{equation}%
where we have set $u_{0}=1$ and $u_{1}=0$. This is a natural generalization
of the well known differential $sl_{2}$-Lax operator ${\cal L}_{2}={\xi _{H}}%
^{2}+u_{2}$ associated to the KdV integrable hierarchy on the torus $T^{2}$
that we will discuss later. Consider the KdV Lax operator that we can write
by virtue of the Miura transformation as
\begin{equation}
\begin{array}{lcl}
{\cal L}({u}_{2}) & = & \xi _{H}^{2}+u_{2}(z,\omega ) \\
& = & (\xi _{H}+\{logH,\phi \})\times (\xi _{H}-\{logH,\phi \})%
\end{array}%
\end{equation}%
where $\phi $ is a Lorentz scalar field. As a result we have
\begin{equation}
u_{2}=-\{logH,\phi \}^{(2)}-\{logH,\phi \}^{2}
\end{equation}%
describing the classical version of the stress energy momentum tensor of
conformal field theory on the torus $T^{2}$. Using bicomplex coordinates, we
can write
\begin{equation}
{\cal T}(z,\omega )\equiv u_{2}(z,\omega )=-\{logH,\phi \}^{(2)}-\{logH,\phi
\}^{2}
\end{equation}%
The conservation for this bianalytic conformal current ${\cal T}(z,\omega )$%
, leads to write the following differential equation
\begin{equation}
\{log{\bar{K}},\{logH,\phi \}\}=e^{2\phi }
\end{equation}%
where ${\bar{K}}=K(\bar{z},\bar{\omega})$ is an arbitrary bianalytic
function of $\bar{z}$ and $\bar{\omega}$ carrying in general an $({\bar{n}}%
_{0},{\bar{m}}_{0})$ $U(1)\times U(1)$ charge. Note also that ${\bar{K}}$ is
not necessarily the complex conjugate of the function $H$ considered
earlier. Our experience with conformal field theory and integrable systems
leads to conclude that the later "second order" differential equation is
nothing but the conformal Liouville like equation of motion on the Torus $%
T^{2}$. This equation of motion is known to appear in this context as a
compatibility relation with the conservation of the stress energy momentum
tensor ${\cal T}(z,\omega )$ namely
\begin{equation}
\{log{\bar{K}},{\cal T}(z,\omega )\}=0
\end{equation}%
or equivalently
\begin{equation}
\{log{\bar{K}},\{logH,\phi \}^{(2)}\}+2\{logH,\phi \}\{log{\bar{K}}%
,\{logH,\phi \}\}=0
\end{equation}

\subsection{The KdV equation on $T^2$}

The KdV-like Lax operator
\begin{equation}
{\cal L}_{KdV}=\xi _{H}^{2}+u_{2}(z,\omega )
\end{equation}%
belongs to the coset space ${{\widehat{\Sigma }}_{2}^{(0,2)}}/{{\widehat{%
\Sigma }}_{2}^{(1,1)}}$. As known from standard references in non-linear
integrable models, we can set by analogy
\begin{equation}
\frac{\partial {\cal L}}{\partial t_{2n+1}}=[({\cal L})_{+}^{\frac{2n+1}{2}},%
{\cal L}]
\end{equation}%
which gives the $n-th$ evolution equation of the KdV-hierarchy. The index +
in eq.(33), stands for the local part of the pseudo-differential operator $%
{\cal L}^{\frac{2n+1}{2}}$ defined as follows ${\cal L}^{\frac{2n+1}{2}}=%
{\cal L}^{\frac{1}{2}}\circ {\cal L}^{n}$ where ${\cal L}^{\frac{1}{2}}$ is
nothing but the half power of the KdV Lax operator. It describes a
pseudo-differential operator weight $|{\cal L}^{\frac{1}{2}}|=(1,1)$. The
non linear pseudo-differential operator ${\cal L}^{\frac{2n+1}{2}}$
describes the $(2n+1)^{th}$ power of ${\cal L}^{\frac{1}{2}}$
\begin{equation}
{\cal L}^{\frac{1}{2}}=\xi _{H}+\frac{1}{2}u_{2}\xi _{H}^{-1}-\frac{1}{4}%
\{logH,u_{2}\}\xi _{H}^{-2}+[\frac{1}{8}\{logH,u_{2}\}^{(2)}-\frac{1}{8}%
u_{2}^{2}]\xi _{H}^{-3}...
\end{equation}%
where the coefficients are explicitly determined by requesting ${\cal L}%
_{2}=({\cal L}^{\frac{1}{2}}\circ {\cal L}^{\frac{1}{2}})$. \newline
Consider special orders of the hierarchy eq(33) parametrized by the index $n$%
. For $n=0$ we get
\begin{equation}
\frac{\partial {\cal L}}{\partial t_{1}}=[({\cal L})_{+}^{\frac{1}{2}},{\cal %
L}]
\end{equation}%
where $({\cal L})_{+}^{\frac{1}{2}}=\xi _{H}=\{logH,.\}$. We show also that
eq.(36) corresponds simply to the chiral wave equation,
\begin{equation}
\frac{\partial u_{2}}{\partial t_{1}}=\{logH,u_{2}\}
\end{equation}%
For $n=1$, we have
\begin{equation}
\frac{\partial {\cal L}}{\partial t_{3}}=[({\cal L})_{+}^{\frac{3}{2}},{\cal %
L}]
\end{equation}%
where $({\cal L}_{+}^{\frac{3}{2}})_{+}$ is explicitly given by
\begin{equation}
({\cal L}^{\frac{3}{2}})_{+}=\xi _{H}^{3}+\frac{3}{2}u_{2}\xi _{H}+\frac{3}{4%
}\{logH,u_{2}\}
\end{equation}%
Injecting this expression into eq.(38) we get a non linear differential
equation giving the evolution of the spin-$2$ conformal current $u_{2}$,
once some easy algebraic manipulations are performed. This is nothing but
the KdV equation on the bidimenaional torus $T^{2}$ given by
\begin{equation}
\frac{\partial u_{2}}{\partial t_{3}}=\frac{3}{2}u_{2}\{logH,u_{2}\}+\frac{1%
}{4}\{logH,u_{2}\}^{(3)}
\end{equation}%
The same computations hold for the other evolution equations.

\section{Conserved quantities}

Actually we know that the KdV equation in diff$(S^{1})$ case is an
integrable equation. This is because its non linear behavior dealing with
solitonic solutions implies the existence of an infinite number of conserved
quantities. The determination of these conserved quantities is well known in
the standard case. We are presently looking for the diff$(T^{2})$ extension
and it's impact on the integrability process. Let's ${\cal Q}[u_{i}]$ be a
conserved quantity for which we assume the following
\begin{equation}
\frac{d{\cal Q}{[u_{i}]}}{dt}=[{\cal Q}[u_{i}],{\cal H}]
\end{equation}%
where ${\cal H}$ is the hamiltonian of the system and ${\cal Q}(u_{i})$
reads in terms of the charge-density as
\begin{equation}
{\cal Q}[u_{i}]=\int dz.d\omega \rho \lbrack u_{i}]
\end{equation}%
For the time-independent charges ${\cal Q}(u_{i})$ we get the following
continuity equation
\begin{equation}
\frac{\partial \rho \lbrack u_{i}]}{\partial t_{k}}+\{log{\tilde{H}}%
,j[u_{k}]\}=0
\end{equation}%
Using the following natural property
\begin{equation}
\frac{1}{n}\{log{\tilde{H}},f^{n}\}=f^{n-1}\{log{\tilde{H}},f\}
\end{equation}%
the KdV equation (40) reads as
\begin{equation}
\begin{array}{lcl}
\frac{\partial u_{2}}{\partial t_{k}} & = & \frac{3}{2}\{log{\tilde{H}},%
\frac{u^{2}}{2}\}+\frac{1}{4}\{log{\tilde{H}},u\}^{(3)} \\
& = & \{log{\tilde{H}},[\frac{3u^{2}}{4}+\frac{1}{4}\{log{\tilde{H}}%
,u\}^{(2)}]\}%
\end{array}%
\end{equation}%
Combined with the continuity equation eq.(44), leads to
\begin{equation}
\begin{array}{lcl}
\rho _{_{0}}[u] & = & u_{2} \\
{j}_{_{0}} & = & \frac{3u^{2}}{4}+\frac{1}{4}\{log{\tilde{H}},u\}^{(2)}%
\end{array}%
\end{equation}%
Thus, a constant of motion can be extracted, namely
\begin{equation}
\begin{array}{lcl}
{\cal Q}_{_{0}}={\cal H}_{_{0}} & = & \int dz.d\omega \rho _{_{0}}[u_{i}] \\
& = & \int dz.d\omega u_{2}%
\end{array}%
\end{equation}%
The next steps concerns the determination of other constant of motion.
Consider once again the KdV equation (40), we obtain
\begin{equation}
\frac{\partial }{\partial t_{k}}(\frac{1}{2}u^{2})=\{log{\tilde{H}},[\frac{%
u^{3}}{3}-\frac{1}{8}\{log{\tilde{H}},u\}^{2}+\frac{1}{4}u\{log{\tilde{H}}%
,u\}^{(2)}]\}
\end{equation}%
Consequently, a second continuity equation can be extracted, that's
\begin{equation}
\frac{\partial \rho \lbrack u_{i}]}{\partial t_{k}}+\{log{\tilde{H}},{j}%
[u_{k}]\}=0
\end{equation}%
with
\begin{equation}
\begin{array}{lcl}
\rho _{_{1}[u]} & = & u_{2} \\
{j}_{_{1}} & = & -\frac{u^{3}}{3}+\frac{1}{8}\{log{\tilde{H}},u\}^{2}-\frac{1%
}{4}\{log{\tilde{H}},u\}^{(2)}%
\end{array}%
\end{equation}%
The second constant of motion is then given by
\begin{equation}
\begin{array}{lcl}
{\cal Q}_{_{1}}={\cal H}_{_{1}} & = & \int dz.d\omega \rho _{_{1}}[u_{i}] \\
& = & \int dz.d\omega (\frac{1}{2}u^{2})%
\end{array}%
\end{equation}

\section{Lax-Pair representation}

\subsection{Lax-Pair of the KdV equation}

It's commonly known that Lax pair operators, once they exist, paly a central
role in proving the integrability. An integrable equation which posses the
Lax representation can be rewritten into the form of Lax equation given by
\begin{equation}
\lbrack {\cal L}=\xi _{H}^{2}+u_{2},{\cal P}+\partial _{t_{KdV}}]=0
\end{equation}%
with $\partial _{t}=\frac{\partial }{\partial t}$ and $t_{KdV}\equiv t_{3}$.
Given a Lax operator ${\cal L}$, the crucial point in the Lax-pair technique
is to find a corresponding operator ${\cal P}$ constrained by eq.(53). This
problem is very difficult to solve in general. However, putting some ansatz
on ${\cal P}$, can help to get a wide class of solutions.\newline
\ \ \ \ \ \ \ \ \ \ \ \ \ \ {\bf Ansatz}: $\ \ \ \ \ \ \ \ \ \ \ \ \ \ \ \ \
\ \ \ \ \ \ \ \ \ \ \ \ \ \ \ \ \ \ \ \ \ \ \ \ \ \ {\cal P}=\xi
_{H}^{r}\circ {\cal L}^{s}+{\cal P^{\prime }}$

This ansatz reduces, in some sense, the problem for ${\cal P}$ to that for $%
{\cal P^{\prime }}$. So, let's consider the KdV Lax operator eq.(32)
corresponding to $r=s=1$, the bracket eq.(53) reduces, after easy
computations, to
\begin{equation}
\lbrack \xi _{H}^{2}+u_{2},{\cal P}^{\prime }]={\partial _{t_{3}}}{u_{2}}%
+u_{2}.\{logH,u_{2}\}+\{logH,u_{2}\}\xi _{H}^{2}
\end{equation}%
Since the l.h.s. of this equation is a term that belongs to the ring ${\cal R%
}$, consistency requires that we should delete the term in $\xi _{H}^{2}$
figuring in it's r.h.s. This can be done if one take the following form for
the operator ${\cal P}^{\prime }$
\begin{equation}
{\cal P}^{\prime }=\frac{1}{2}u_{2}{\xi _{H}}-\frac{1}{4}\{logH,u_{2}\}
\end{equation}
Plugging these expressions into the Lax equation (53), one obtain an
expression for the operator ${\cal P}$ similar to that of $({\cal L}^{\frac{3%
}{2}})_{+}$ eq.(39), in fact we get
\begin{equation}
{\cal P}={\xi _{H}^{3}}+\frac{3}{4}\{logH,u_{2}\}+\frac{3}{2}u_{2}\xi _{H}
\end{equation}%
With this expression of the operator ${\cal P}$, the Lax equation eq.(53)
leads to recover the same form of the KdV equation established in eq.(40).
We have then the explicit form of the Lax-pair $({\cal L,P})$, associated to
the KdV equation on the torus $T^{2}$.

\subsubsection{Lax-pair of the Burgers Equation}

Our interest in the Burgers equation comes from the several important
properties that are exhibited in the diff$(S^{1})$ ca se. Let's recall for
instance that in the standard pseudo-differential operator's formalism, this
equation is associated to the following ${L}$-operator
\begin{equation}
L_{Burg}=\partial _{x}+u_{1}(x,t)
\end{equation}%
where the function $u_{1}$ is of conformal spin one. Note that in the
complex language where $z=x+it$ and $\bar{z}=x-it$, one can write $u=u(z,%
\bar{z})$ and show that under a conformal change of coordinate $z\rightarrow
{\tilde{z}}=f(z)$ the $u_{1}$ currents transforms as an object of conformal
spin $1$. Now, we are ready to look for the diff$(T^{2})$ version of the
Burgers equation associated to
\begin{equation}
{\cal L}_{Burg}=\xi _{H}+u_{1}(z,\omega ).
\end{equation}
This is a local differential operator of the generalized $n$-KdV hierarchy's
family $(n=1)$, that we can interpret as been the result of a truncation of
a pseudo differential operator of KP-hierarchy type
\begin{equation}
{\cal L}_{KP}=\xi _{H}+u_{1}(z,\omega )+u_{2}(z,\omega )\circ \xi
_{H}^{-1}+u_{3}(z,\omega )\circ \xi _{H}^{-2}+...,
\end{equation}%
of the space ${\widehat{\Sigma }}_{1}^{(-\infty ,1)}$. The local truncation
is simply given by
\begin{equation}
{\widehat{\Sigma }}_{1}^{(-\infty ,1)}\rightarrow {\widehat{\Sigma }}%
_{1}^{(0,1)}\equiv \lbrack {\widehat{\Sigma }}_{1}^{(-\infty ,1)}]_{+}\equiv
{\widehat{\Sigma }}_{1}^{(-\infty ,1)}/{\widehat{\Sigma }}_{1}^{(-\infty
,-1)},
\end{equation}%
or equivalently
\begin{equation}
{\cal L}_{1}(u_{i})=\xi _{H}+\Sigma _{i=0}^{\infty }u_{i}\xi
_{H}^{1-i}\rightarrow \xi _{H}+u_{1}\equiv \lbrack {\cal L}_{1}(u_{i})]_{+},
\end{equation}
\ The diff$(T^{2})$-Burgers equation is said to have the Lax representation
if there exists a suitable pair of operators $({\cal L},{\cal P})$ so that
the commutation Lax equation
\begin{equation}
\lbrack {\cal L},{\cal P}+\partial _{t_{Burg}}]=0,
\end{equation}%
with $t_{Burg}\equiv t_{2}$. To generate the Lax-pair associated to the diff$%
(T^{2})$-version of the Burgers equation one have to consider the following
ansatz for the operator ${\cal P}$
\begin{equation}
{\cal P}=\xi _{H}\circ {\cal L}+{\cal P}^{\prime },
\end{equation}%
or
\begin{equation}
{\cal P}=\xi _{H}^{2}+u_{1}\xi _{H}+\{logH,u_{1}\}+{\cal P}^{\prime }.
\end{equation}%
Performing straightforward computations, the diff$(T^{2})$ Burgers Lax
equation, reduces then to
\begin{equation}
\lbrack {\cal L}_{Burg}=\xi _{H}+u,{\cal P}^{\prime }]=\{logH,u_{1}\}\circ
\xi _{H}+(u_{1}\{logH,u_{1}\}-\frac{1}{2}\{logH,u_{1}\}^{(2)}+\partial
_{t_{Burg}}u_{1})
\end{equation}%
Next, one have to take the following ansatz for the operator ${\cal P}%
^{\prime }$
\begin{equation}
{\cal P}^{\prime }=A\xi _{H}+B,
\end{equation}%
where $A$ and $B$ are arbitrary functions for the moment. With this new
ansatz for ${\cal P}^{\prime }$, we have
\begin{equation}
\begin{array}{lcl}
\lbrack \xi _{H}+u_{1},{\cal P}^{\prime }]= & - & (A\{logH,u_{1}\}+\frac{1}{2%
}\{logH,A\}^{(2)}+\frac{1}{2}[u,\{logH,A\}]+\{logH,B\}+[u,B]) \\
&  &  \\
& + & (\{logH,A\}+[u_{1},A])\xi _{H}%
\end{array}%
\end{equation}%
\newline
Identifying eqs.(68) and (70) leads to the following constraints equations
\begin{equation}
\{logH,u_{1}\}=\{logH,A\}+[u_{1},A]
\end{equation}%
and
\begin{equation}
(u_{1}+A)\{logH,u_{1}\}+\partial _{t_{Burg}}u_{1}=\{logH,B\}+[u_{1},B]+\frac{%
1}{2}[\{logH,A\},u_{1}]+\frac{1}{2}[\{logH,(u_{1}-A)\}^{(2)}
\end{equation}%
A natural solution of the first constraint equation (71) is $A=u_{1}$. This
implies a reduction of eq.(72) to
\begin{equation}
2u_{1}\{logH,u_{1}\}+\partial _{t_{Burg}}u_{1}=\{logH,B\}+[u_{1},B]
\end{equation}
Since $B$ is an object of conformal weight 2, We consider the following
solution for eq.(73)
\begin{equation}
B=\alpha \{logH,u_{1}\}+\beta u^{2}
\end{equation}%
with $\alpha $ and $\beta $ are arbitrary coefficient numbers. Injecting
this expression into eq.(73) gives the final expression of the diff$(T^{2})$%
-Burgers equation namely
\begin{equation}
{\partial _{t_{Burg}}{u_{1}}}+2(1-\beta )u_{1}\{logH,u_{1}\}-\alpha
\{logH,u_{1}\}^{(2)}=0
\end{equation}%
The associated Lax-pair is given by
\begin{equation}
{\cal L}_{Burg}=\xi _{H}+u_{1}
\end{equation}%
and
\begin{equation}
{\cal P}_{Burg}=\xi _{H}^{2}+2u_{1}\xi _{H}+\beta u_{1}^{2}+\alpha
\{logH,u_{1}\}
\end{equation}

\subsection{\bf Burgers-KdV mapping}

This subsection will be devoted to another significant aspect of integrable
models in diff$(T^{2})$ framework. The principal focus, for the moment, is
on the models discussed previously namely the KdV and Burgers systems.
Previously, we discussed the integrability of these two nonlinear systems
and we noted that they are indeed integrable and this property is due to the
existence of definite Lax pair operators $({\cal {L,P}})_{Burg}$ for each of
the two models. Such existence implies the linearization of the models
automatically. A crucial question which arises now is to know if there is a
possibility to establish a mapping between the two Systems. The idea to
connect the two models is originated from the fact that integrability for
the KdV system is something natural due to conformal symmetry. We think that
the strong backgrounds of conformal symmetry on $T^{2}$ can help to built
integrability of the Burgers systems if one know how to establish such a
connection.\newline
{\bf Proposition 1:}\newline
Consider the Burgers operator ${\cal L}_{Burg}(u_{1})=\xi _{H}+u_{1}\in {%
\widehat{\Sigma }}_{1}^{(0,1)}$, for any given KdV operator ${\cal L}%
_{KdV}(u_{2})=\xi _{H}^{2}+u_{2}$ belongings to the space ${\widehat{\Sigma }%
}_{2}^{(0,2)}/{\widehat{\Sigma }}_{2}^{(1,1)}$, one can define the following
mapping
\begin{equation}
{\widehat{\Sigma }}_{1}^{(0,1)}\hookrightarrow {\widehat{\Sigma }}%
_{2}^{(0,2)}/{\widehat{\Sigma }}_{2}^{(1,1)},
\end{equation}%
in such away that
\begin{equation}
{\cal L}_{Burg}(u_{1})\rightarrow {\cal L}_{KdV}(u_{2})\equiv {\cal L}%
_{Burg}(u_{1})\otimes {\cal L}_{Burg}(-u_{1}).
\end{equation}%
What we are assuming in this proposition is a strong constraint leading to
connect the two spaces. This constraint is also equivalent to set
\begin{equation}
{\widehat{\Sigma }}_{2}^{(0,2)}/{\widehat{\Sigma }}_{2}^{(1,1)}\equiv {%
\widehat{\Sigma }}_{1}^{(0,1)}\otimes {\widehat{\Sigma }}_{1}^{(0,1)}
\end{equation}%
Next we are interested in exploring the crucial key behind the previous
proposition, we underline then that this mapping is easy to highlight
through the Miura transformation
\begin{equation}
{\cal L}_{KdV}=\xi _{H}^{2}+u_{2}=(\xi _{H}+u_{1})\circ (\xi _{H}-u_{1})
\end{equation}%
giving rise to
\begin{equation}
u_{2}=-u_{1}^{2}-\{logH,u_{1}\}
\end{equation}%
The {\em proposition 1} can have a complete and consistent significance only
if one manages to establish a connection between the differential equations
associated to the two systems. At this stage, note that besides the
principal difference due to conformal spin, we stress that the two nonlinear
evolutions equations of KdV
\begin{equation}
{\partial _{t_{3}}u_{2}}=\frac{3}{2}u_{2}\{logH,u_{2}\}+\frac{1}{4}%
\{logH,u_{2}\}^{(3)}
\end{equation}%
and of Burgers
\begin{equation}
{\partial _{t_{Burg}}{u_{1}}}+2(1-\beta )u_{1}\{logH,u_{1}\}-\alpha
\{logH,u_{1}\}^{(2)}=0
\end{equation}%
are distinct by a remarkable fact that is the KdV flow $t_{KdV}\equiv t_{3}$
and the Burgers one $t_{Burg}\equiv t_{2}$ have different conformal weights:
$[t_{KdV}]=(-3,-3)$ whereas $[t_{Burg}]=(-2,-2)$. \newline
Now, we are constrained to circumvent the effect of proper aspects specific
to both the equations and consider the following second property: \newline
{\bf Proposition 2:}\newline
By virtue of the Burgers-KdV mapping and dimensional arguments, the
associated flow are related through the following ansatz
\begin{equation}
(\partial _{t_{_{Burg.}}}{\bullet })\hookrightarrow (\partial _{t_{_{KdV}}}{%
\bullet })\equiv \partial _{t_{_{Burg.}}}.\{logK,{\bullet }\}+\eta \{logK,{%
\bullet }\}^{(3)}
\end{equation}%
acting on arbitrary function ${\cal F}$ in the following way
\begin{equation}
\partial _{t_{_{Burg.}}}{\cal F}\hookrightarrow \partial _{t_{_{KdV}}}{\cal F%
}\equiv \{logK,(\partial _{t_{_{Burg.}}}{\cal F})\}+\eta \{logK,{\cal F}%
\}^{(3)}
\end{equation}%
for an arbitrary parameter $\alpha $. With respect to the assumption
eq.(85), relating the two evolution derivatives $\partial _{t_{_{Burg.}}}$
and $\partial _{t_{_{KdV}}}$ , one should expect some strong constraint on
the Burgers differential equation (84). Using {\em proposition 2}, we have
to identify the following three differential equations
\begin{equation}
\begin{array}{lcl}
{\partial _{t_{3}}u_{2}} & = & \frac{3}{2}u_{2}\{logH,u_{2}\}+\frac{1}{4}%
\{logH,u_{2}\}^{(3)} \\
&  &  \\
& = & -2u_{1}\partial _{t_{3}}u_{1}-\partial _{t_{3}}\{logH,u_{1}\}, \\
&  &  \\
& = & \partial _{t_{2}}\{logH,u_{2}\}+\eta \{logH,u_{2}\}^{(3)}.%
\end{array}%
\end{equation}%
Setting for a matter of simplicity the Burgers equation as $\partial
_{t_{2}}u_{1}=au_{1}\{logH,u_{1}\}+b\{logH,u_{1}\}^{(2)}$ with $a=2(\beta
-1) $ and $b=\alpha $, and performing explicit computation, rising from the
identification of the previous system of equations (87), we find\newline
\begin{equation}
\begin{array}{lcl}
\partial _{t_{3}}u_{2} & = & 3u_{1}^{3}\{logH,u_{1}\}+3\{logH,u_{1}%
\}^{2}u_{1}+\frac{3}{2}\{logH,u_{1}\}^{(2)}u_{1}^{2}-\frac{1}{2}%
\{logH,u_{2}\}^{(3)}u_{1} \\
&  &  \\
&  & -\frac{1}{4}\{logH,u_{1}\}^{(4)} \\
&  &  \\
& = & -2a\{logH,u_{1}\}^{2}u_{1}-3a\{logH,u_{1}\}^{(2)}u_{1}^{\prime
}-2a\{logH,u_{1}\}^{2}\{logH,u_{1}\}^{(2)} \\
&  &  \\
&  & -(b+\eta )\{logH,u_{1}\}^{(3)}-2(\eta +\frac{a}{2}+b)u_{1}\{logH,u_{1}%
\}^{(3)} \\
&  &  \\
& = & -4a\{logH,u_{1}\}^{2}u_{1}-2(b+3\eta +\frac{3a}{2})\{logH,u_{1}%
\}^{(2)}\{logH,u_{1}\} \\
&  &  \\
&  & -2(b+\eta +\frac{a}{2})\{logH,u_{1}\}^{(3)}u_{1}-2a\{logH,u_{1}%
\}^{(2)}u_{1}^{2}-(b+\eta )\{logH,u_{1}\}^{(4)}%
\end{array}%
\end{equation}%
These expressions, once are simplified, lead to a strong constraint on the
Burgers equation. Performing straightforward computations one shows that
\begin{equation}
\{logH,u_{1}\}^{(k)}\sim u_{1}^{k+1},\hspace{1.5cm}1\leq k\leq 4
\end{equation}%
which means that $\{logH,u_{1}\}\sim u_{1}^{2}$, $\{logH,u_{1}\}^{(2)}\sim
u_{1}^{3}$ and so one. \newline
Putting these constraint equations into the Burgers equation (84) one obtain
the following differential equation
\begin{equation}
{\partial _{t_{Burg}}}u_{1}\sim \{logH,u_{1}\}^{(2)}
\end{equation}%
This is an impressing result since the mapping between the flow of KdV and
Burgers nonlinear differential equations affects the Burgers equation as
follows
\begin{equation}
{\partial _{t_{Burg}}{u_{1}}}\sim
(...)u_{1}\{logH,u_{1}\}+(...)\{logH,u_{1}\}^{(2)}\hookrightarrow {\partial
_{t_{Burg}}{u_{1}}}\sim \{logH,u_{1}\}^{(2)}
\end{equation}%
This is also equivalent to argue that the proposed mapping induces a
cancelation of the non linear term $\sim (...)u_{1}\{logH,u_{1}\}$
responsible of solitonic character at the level of the Burgers equation. We
guess that a hidden extended $2d$-conformal symmetry is behind the
linearizability property induced by the Burgers-KdV mapping. This is because
the conformal symmetry in the framework of KdV hierarchy is related in
general to the $sl_{n}$-symmetry. In fact, we have to remark that the
Burgers $u_{1}$-current issued from the Miura like equation (79) can be
identified with the Liouville Lorentz scalar field $\phi $ as follows $%
u_{1}\equiv \{logK,\phi \}$ describing the derivative of the Liouville
Lorentz scalar field while the KdV potential $u_{2}$ satisfying eq.(83) can
be then identified with the conformal current ${\cal T}$ given by eq.(28).

\section{Concluding Remarks}

We presented in this paper some important aspects of integrable KdV
hierarchies dealing with higher conformal spin symmetries on the
bidimensional torus $T^{2}$. These symmetries, generalizing the Frappat et
al. conformal symmetries by adding currents of conformal spin $(3,3)$ in a
non standard way, are also shown to be derived, in their semi-classical
form, from the GD bracket \cite{sed 07}. \newline
Note that KdV hierarchies on diff$(T^{2})$ exhibits many remarkable
features. The first one concerns the introduction of new kind of derivatives
taking the following form $\xi _{H}\equiv \{logH,.\}=\partial
_{z}logH\partial _{\omega }-\partial _{\omega }logH\partial _{z}$ for
arbitrary bianalytic function $H(z,\omega )$. Besides the above established
results, we tried also to understand much more the meaning of integrability
of nonlinear systems on $T^{2}$. The principal focus was on the KdV and
Burgers systems. A first step was to derive these two equations using the
above systematic algebraic formulation in the context of Lax-pair building
program. Concerning the derived KdV system, this is an integrable model due
to the existence of a Lax pair operators $({\cal L}_{_{KdV}},{\cal P}%
_{_{KdV}})$. This existence is an important indication of integrability, but
we guess that the realistic source of integrability of this model is the
underlying conformal symmetry. For the Burgers system, to check its
integrability we proceeded to an explicit derivation of the Lax pair
operators $({\cal L}_{_{Bur}},{\cal P}_{_{Bur}})$ giving rise to the
following differential equation
\begin{equation}
2u_{1}\{logH,u_{1}\}+\partial _{t_{Burg}}u_{1}=\{logH,B\}+[u_{1},B].
\end{equation}%
Solving this equation, we get the explicit form of the requested Lax
operator.\newline
Concerning the possibility to establish a correspondence between the KdV and
the Burgers systems, actually, we succeeded to build a mapping from the
Burgers system to the KdV one. The main lines of this mapping deals with the
following ansatz
\begin{equation}
\partial _{t_{_{Burg.}}}{\cal F}\hookrightarrow \partial _{t_{_{KdV}}}{\cal F%
}\equiv \{logK,(\partial _{t_{_{Burg.}}}{\cal F})\}+\eta \{logK,{\cal F}%
\}^{(3)}
\end{equation}%
for an arbitrary parameter $\eta $. \newpage {\small {\ {\bf Acknowledgments}%
\newline
I would like to thank the Abdus Salam International Center for
Theoretical Physics (ICTP) for hospitality. I present special thanks
to the high energy section and to its head Seif Randjbar-Daemi. Best
thanks are presented to the office of associates for the invitation
and support. I acknowledge the contribution of OEA-ICTP in the
context of NET-62 and thank K.S. Narain and E.H. Saidi for useful
conversations.} }

\small{

}
\end{document}